# Reaching the Shadowed Equator: Propagation of Radio Emissions in Planetary Magnetospheres

***Siyuan Wu and Daniel K. Whiter***

*Abstract* – The dominant natural radio emissions in the magnetospheres of Earth, Saturn, and Jupiter are primarily generated by the cyclotron maser instability along polar auroral magnetic field lines. Observations show that these emissions are mainly detected at mid-high latitudes because equatorial regions are generally considered radio "shadow zones," where dense plasma structures inhibit their detection. However, recent observations at Saturn and Earth reveal a widespread presence of radio emissions within nominal "shadow regions," where direct propagation was previously thought to be blocked. To account for these detections, several ray-propagation scenarios have been proposed that enable radio waves to access regions traditionally considered observationally inaccessible. Here, we summarize these propagation mechanisms and demonstrate how radio emissions detected within shadow regions can provide valuable diagnostics of magnetospheric structure and dynamics.

## 1. Introduction

The dominant planetary radio emissions observed throughout the solar system are closely associated with auroral processes and are widely attributed to the cyclotron maser instability (CMI) [1]. CMI is a relativistic cyclotron resonance between electromagnetic waves and energetic electrons that generates intense radio emissions near the local electron cyclotron frequency. Examples include auroral kilometric radiation (AKR) at Earth [2], decametric, hectometric, and broadband kilometric (bKOM) emissions at Jupiter [3, 4], and Saturn kilometric radiation (SKR; Figure 1) [5]. These emissions originate along auroral magnetic field lines and represent the dominant radio emission process in magnetized planetary environments. Because they are highly sensitive to magnetospheric activity, they provide powerful remote diagnostics of solar wind–magnetosphere interactions. For example, AKR is strongly correlated with terrestrial substorms [6, 7], SKR responds to solar-wind compressions and internal injection processes [8, 9], and bKOM emissions are modulated by solar-wind conditions [10].

Although the generation mechanism of CMI-driven radio emissions is well established [4, 11, 12], their detectability strongly depends on propagation effects controlled by plasma density structures and magnetic-field geometry. The visibility of auroral radio emissions is therefore constrained by beaming geometry and observer location, requiring propagation effects to be considered when interpreting observations. Recent studies [13, 14] have revealed unexpected propagation paths for both SKR and AKR, showing that auroral radio emissions can access equatorial regions previously considered observationally inaccessible. These include shadow regions associated with dense plasma structures, such as the plasma torus at Saturn and Jupiter and the plasmasphere at Earth, as well as extended equatorial regions beyond these structures. Several non-trivial ray-propagation scenarios have been proposed to explain these observations, challenging the conventional view that equatorial regions are effectively shielded from auroral radio emissions. Here, we summarize these propagation results and discuss their implications for interpreting planetary radio observations.

Manuscript received 13 April 2026. This work was supported by the Science and Technology Facilities Council (STFC) of the UK under Grant ST/Y002121/1.

Siyuan Wu and Daniel K. Whiter are with University of Southampton, University Road, Southampton, Hampshire SO17 1BJ, UK; e-mail: wusiyuan826@gmail.com, D.Whiter@soton.ac.uk.

## 2. Observation Examples at Saturn

Figure 2 presents SKR observations indicating strong refraction and reflection near the Enceladus plasma torus (EPT). Figures 2 (a) and (b) show pronounced variations in the low-frequency cutoff of SKR as Cassini approaches the EPT, implying that SKR emissions cannot directly propagate into the dense plasma torus and are instead reflected or redirected. The R-X mode SKR exhibits a low-frequency cutoff at $f_x = \frac{1}{2} f_{ce} + \sqrt{f_{pe}^2 + \frac{f_{ce}^2}{4}}$, where $f_{ce}$ and $f_{pe}$ are the electron cyclotron and plasma frequencies, respectively. Because $f_{ce} \ll f_{pe}$ near the torus boundary, $f_x$ approaches the local upper-hybrid frequency $f_{uh} = \sqrt{f_{pe}^2 + f_{ce}^2}$, which is identified by the narrow cutoff "line" marked by the white labels. This cutoff frequency increases with the elevated plasma density at the torus boundary, leading to effective reflection of low-frequency SKR. The second example shown in Figures 2 (c) and (d) exhibits similar behavior, with SKR emissions directly cut off at $f_{uh}$, indicating that the plasma torus boundary acts as an effective propagation barrier for R-X mode SKR. These cutoff emissions display reduced intensity and depolarization (Figure 2d), consistent with strong wave–plasma interactions at the torus boundary.

Nevertheless, statistical results [14] further demonstrate that SKR emissions are ubiquitously observed within the outer part of the plasma torus and in the equatorial region outside the torus, albeit with much lower occurrence rates. This indicates that SKR emissions across all wave frequencies can penetrate or bypass the dense plasma torus, despite the strong cutoff and reflection observed in individual events as shown in Figure 2, motivating further investigation of nontrivial propagation paths. A similar effect is also observed at Earth [13] and will be discussed later.

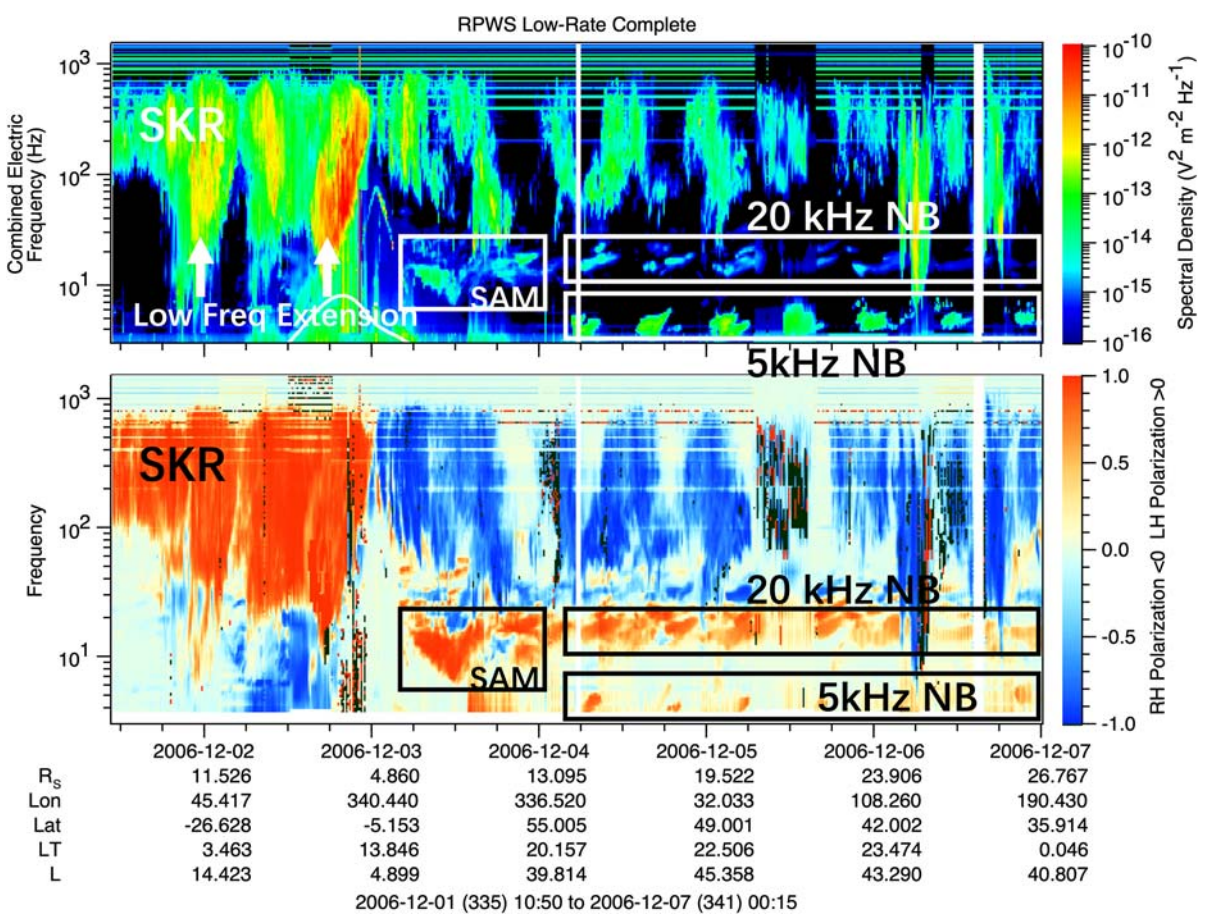

Figure 1. Examples of Saturn radio emissions observed by the Cassini Radio and Plasma Waves instrument. (a) Radio wave electric field spectrogram. The white line indicates the local electron cyclotron frequency along Cassini's orbit. (b) Normalized Stokes V parameter giving the wave polarization features of the wave, with red indicating the left-hand polarized wave and blue marking the right-hand polarized wave.

## 3. Simulation of Ray Paths

CMI-generated emissions are theoretically predicted to be emitted at highly oblique angles relative to the background magnetic field in the source region, as illustrated in Figure 3a. At Saturn, these beaming directions point preferentially toward the plasma torus, wheras at Earth they point toward the plasmasphere. Consequently, an equatorial shadow zone (ESZ) is formed within the intrinsic emission beaming as shown in Figure 3a. Moreover, the elevated plasma density inside the plasma torus further inhibits R-X mode propagation through the cutoff condition at $f_x$, giving rise to an additional R-X forbidden zone, as shown in Figure 3b. Ray-tracing simulations (Figure 3c), performed using the plasma density model and a dipole magnetic field at Saturn [14], demonstrate that high-frequency SKR can traverse the plasma torus with little hindrance. In contrast, emissions below approximately 100 kHz undergo strong refraction and reflection at the torus boundary, causing rays to bend toward higher latitudes. This behavior is consistent with the case observations shown in Figure 2. Together, the ESZ and FZ produce an extended shadowed region that extends beyond the plasma torus, as indicated by the shaded area in Figure 3c.

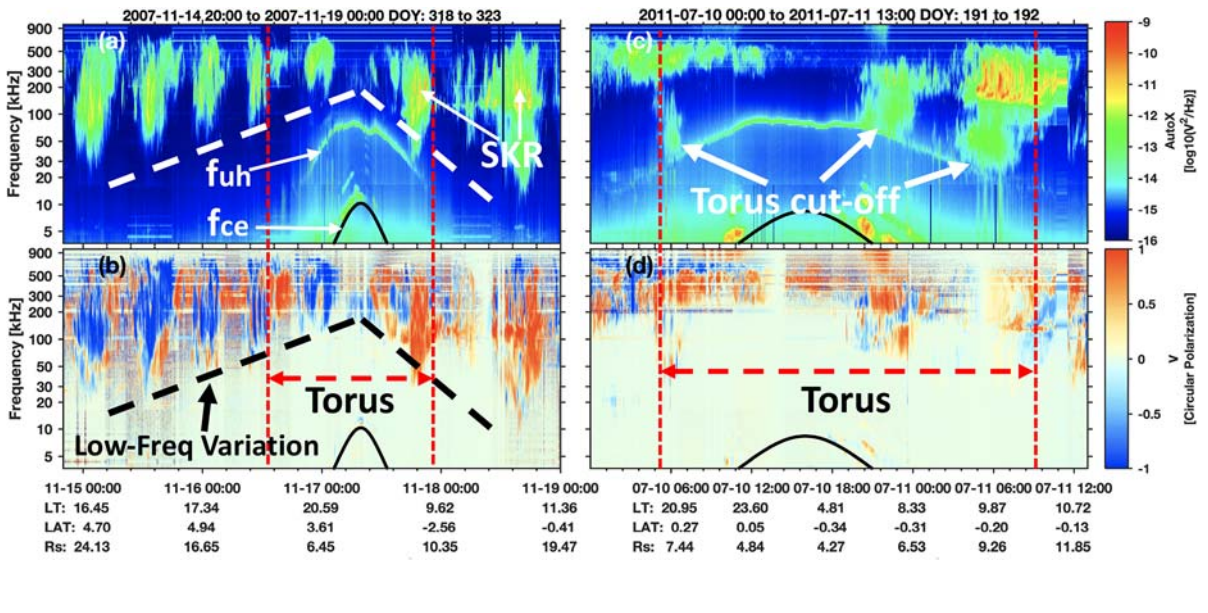

Figure 2. Observations of Saturn Kilometric Radiation near the Enceladus plasma torus (EPT). (a) Shows the electric-field wave spectrogram. (b) Displays the circular polarization. Vertical red lines indicate the approximate location of the EPT. The white and black dashed lines highlight low-frequency variations (c) and (d) Present an additional observation in the same format as panels (a) through (b). Figure modified from [14].

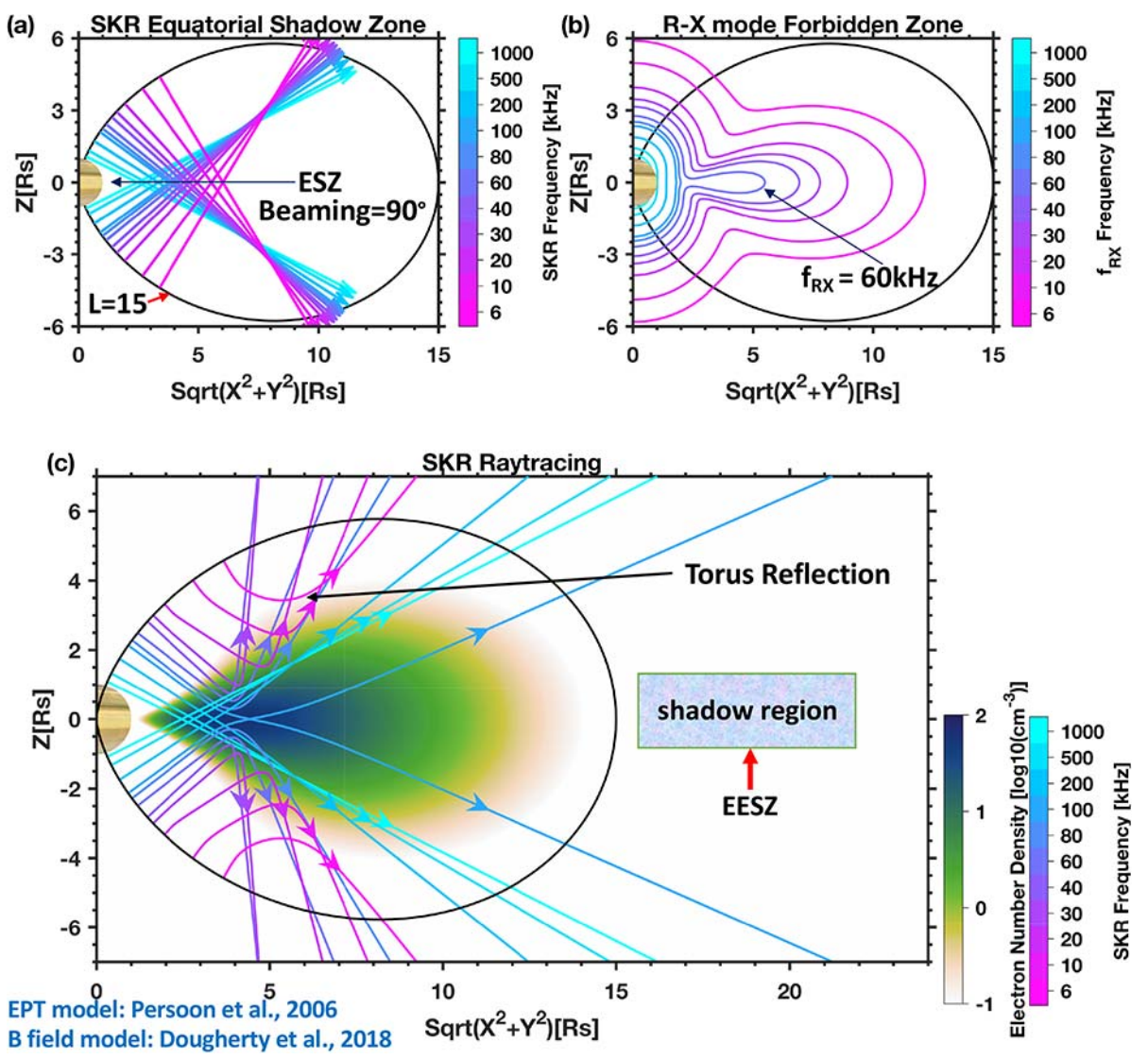

Figure 3. Merging of the equatorial shadow zone (ESZ) and R-X mode forbidden zone (FZ). (a) Displays the beaming pattern of SKR at various frequencies. The beaming angle is chosen to be 90°. The source locations are along the L = 15 magnetic field line. (b) Illustrates the contours of $f_x$ at different frequencies, calculated using the magnetic field model and plasma torus model at Saturn. These $f_x$ contours define the boundaries of the R-X mode FZ. Panel (c) presents the ray-tracing calculation of SKR propagation. The plasma torus model, with electron density indicated by the background color code, is presented. Figure modified from [14].

However, this extended shadow zone contradicts statistical observations as mentioned above, which show a non-negligible occurrence of SKR within this region [14]. This discrepancy suggests that certain propagation paths are not captured by the current ray-tracing framework. A similar issue has recently been identified for terrestrial AKR [13], where emissions are also observed inside the nominal shadow region, implying unresolved or unconventional propagation paths.

## 4. Ray Paths to the Shadow Region

Understanding how radio waves generated along polar auroral field lines can access equatorial "shadow regions," particularly beyond dense plasma structures, such as the plasma torus or plasmasphere, remains a nontrivial problem. A common requirement for shadow-region access is a reduction in local electron density within otherwise-shielding plasma regions. Three broad classes of scenarios have been proposed to enable such propagation: (1) spatial density variations of the plasma torus or plasmasphere, including large-scale gaps or longitudinal asymmetries [14]; (2) temporal density variations, such as rapid contraction or expansion driven by enhanced convection or corotation electric fields [15]; and (3) localized fine-scale density depletions, for example field-aligned low-density ducts [13].

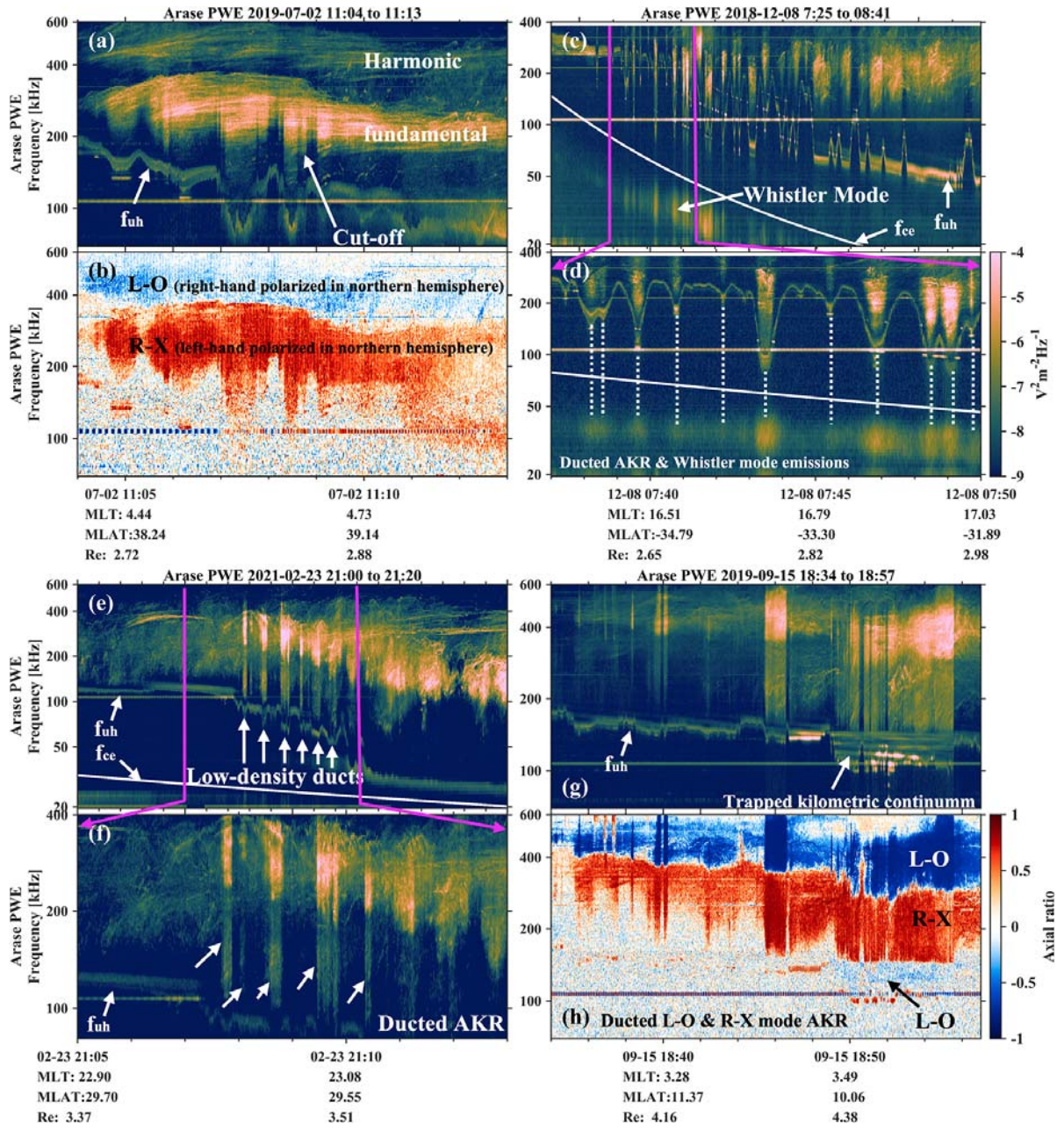

Figure 4. AKR observations near the plasmasphere. (a) Arase PWE wave electric field spectrogram on July 2, 20219 (11:04–11:13), showing enhanced line emissions at the upper-hybrid frequency $f_{uh}$ (white arrow). (b) Axial ratio (circular polarization) over the same interval as (a), with right-hand and left-hand polarized waves shown in blue and red, respectively. (c) Wave electric field spectrogram on December 8, 2018 (07:25–08:41); the white line denotes the local $f_{ce}$. (d) Enlarged view of the spectrogram from 07:37 to 07:50. (e) Wave electric field spectrogram on 2021-02-23 (21:00–21:20). (f) Enlarged view from 21:05 to 21:13 with a different intensity scale. (g) Arase PWE wave electric field spectrogram on September 15, 20219 (18:34–18:57). (h) Corresponding axial ratio spectrogram. Figure modified from [13].

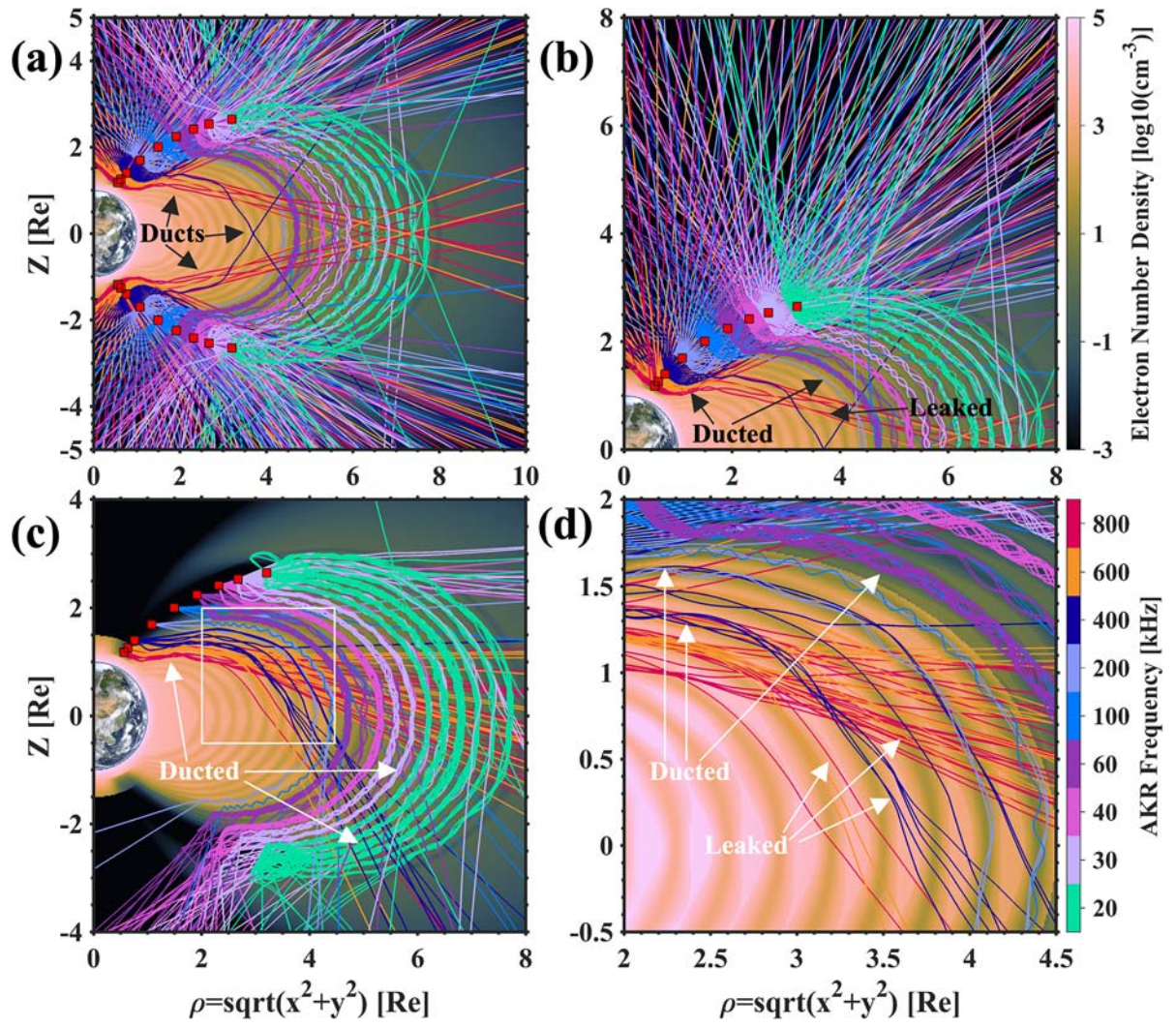

Figure 5. Ray-tracing simulations of AKR propagation. (a) AKR rays in the magnetic meridian plane carried out using the plasmasphere model with the low-density duct structures. (b) Zoomed-in view of the northern hemisphere of (a). (c) Ducted AKR ray paths in the magnetic meridian plane, launched only in the northern hemisphere. (d) Zoomed-in view of (c), as indicated by the white box in (c). Figure modified from [13].

For SKR propagation, scenario (1) appears most plausible. Observations indicate significant longitudinal variability in Saturn's plasma disk [16], suggesting the presence of large-scale density asymmetries, although current measurements remain limited [14]. Such spatial variations could provide pathways for SKR to penetrate regions otherwise shielded by dense plasma. In contrast, recent AKR observations strongly support scenario (3). As shown in Figure 4, AKR frequently exhibits ducted propagation signatures, with emissions modulated by the local $f_{uh}$, indicating confinement within localized density depletions. Ray-tracing results incorporating terrestrial plasma density and magnetic field parameters (Figure 5) demonstrate that AKR can be efficiently guided along low-density ducts and subsequently leak into equatorial shadow regions outside the plasmasphere.

Scenario (2) may, in principle, operate for both AKR and SKR, but its applicability is limited. Plasmaspheric contraction or expansion produces density enhancements either within the compressed inner region or at the expanding boundary, thereby restricting wave propagation over specific frequency ranges. For example, compression leads to enhanced densities on the planetward side, which preferentially suppress higher-frequency emissions while allowing lower-frequency waves to propagate toward the anti-planetward direction. As a result, access to shadow regions is permitted only under limited conditions.

An additional indirect pathway involves reflection at the magnetosheath when the local electron density becomes sufficiently high, a process expected to affect primarily low-frequency emissions. A previous study showed that Saturnian narrowband emissions near 5 kHz can be blocked by the magnetosheath, whereas emissions near 20 kHz can propagate through it [17]. By analogy, partial reflection of SKR may occur only during extreme solar wind compression, when magnetosheath densities are unusually high. A similar mechanism has also been proposed for AKR propagation at Earth [13].

## 5. Emissions Inside the Shadow Region

Emissions detected inside the shadow region arise from specific ray-propagation mechanisms and therefore provide a powerful remote diagnostic of plasma dynamics within the plasma torus or plasmasphere. At Saturn, shadow-region SKR observations can probe plasma torus structure and variability, while at Earth, AKR observations can diagnose low-density ducts within the plasmasphere.

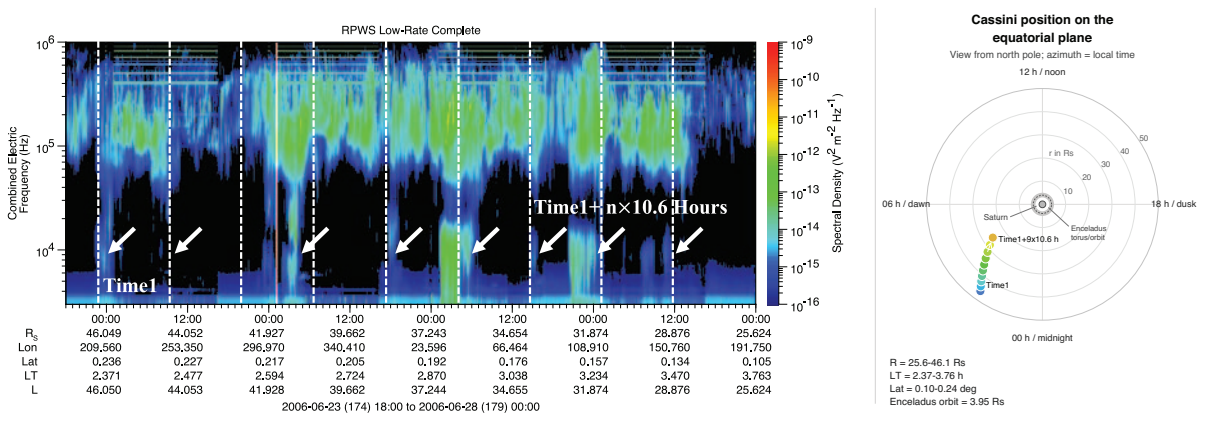

Figure 6. Periodic leakage of SKR emissions observed by Cassini inside the shadow region. The vertical white dashed lines mark successive times given by Time1 + n × 10.6 h, where 10.6 h is approximately Saturn's rotation period.

An example is shown in Figure 6. During this interval, Cassini was located inside the equatorial shadow region (latitude $< 1°$) at radial distances beyond 25 Rs. Although the main SKR band above 100 kHz is continuously observed, emissions below approximately 60 kHz appear weaker, smoother, and detached from the higher-frequency SKR. These low-frequency emissions, commonly referred to as caterpillar emissions because of their characteristic appearance in dynamic spectra [18], are not expected to be visible in this region unless plasma torus variations locally reduce the electron density and allow radio waves to penetrate the shadow region.

The caterpillar emissions exhibit a quasi-periodic recurrence with a period of approximately 10.6 hours, close to Saturn's rotation period, as indicated by the white lines in Figure 6. This suggests that the emissions originate from the same source region and are observed once per planetary rotation as Cassini repeatedly intersects the corresponding ray paths. Their persistence over at least 3 days implies a long-lived density depletion within the plasma torus. Variations in emission intensity further suggest changes in the degree of depletion, with stronger emissions corresponding to weaker refraction and lower plasma densities.

An alternative explanation is a periodic modulation of plasma torus density that temporarily permits radio waves to propagate through the dense plasma region. However, such a scenario would require the plasma density itself to vary with Saturn's 10.6-hour rotation period, for which no observational evidence currently exists. Moreover, because the plasma torus is primarily supplied by Enceladus (orbital period 32.9 hours), a global density modulation synchronized with Saturn's rotation is not expected. In contrast, longitudinal density asymmetries have been reported [16]. Therefore, a long-lived corotating low-density region embedded within the plasma torus remains the most plausible explanation for the observed periodic emissions.

## 6. Conclusion

In this work, we summarized recently proposed ray-propagation pathways that allow planetary radio emissions to access equatorial shadow regions and presented an example demonstrating how such observations can be used to infer the plasma torus structure and variability. Although this example is intended only as an illustration rather than a comprehensive analysis, it highlights the potential of shadow-region radio observations as a complementary diagnostic of magnetospheric plasma conditions.

One aspect not discussed in detail here is the scattering of radio waves by anisotropic plasma density irregularities [19–21], particularly within dense plasma environments such as the terrestrial plasmasphere and Saturnian plasma torus. Previous studies have shown that scattering can significantly modify radio-wave propagation in the Earth's magnetosheath [22] and the solar corona [23]. However, scattering is a cumulative propagation effect, and its contribution over the relatively short propagation distances considered here remains difficult to quantify [24]. More importantly, scattering alone cannot overcome the propagation barrier imposed by dense plasma regions; the density ducts or localized density depletions required for wave penetration must already exist. Therefore, scattering is unlikely to alter the main conclusions of this study, although its quantitative effects warrant future investigation.

## 7. Acknowledgments

Siyuan Wu would like to acknowledge the coauthors of references [13, 14, 17] for their contributions to the results summarized in this work.